\documentclass[ ]{ceurart}

\usepackage{listings}
\usepackage{amsmath}
\usepackage{multirow}
\usepackage{tabularx}
\usepackage{makecell}

\usepackage{booktabs}
\usepackage{array}
\usepackage{url}
\usepackage{subcaption}

\newcolumntype{P}[1]{>{\raggedright\arraybackslash}p{#1}}
\newcolumntype{Y}{>{\raggedright\arraybackslash}X}

\begin{document}

\copyrightyear{2025}
\copyrightclause{Copyright for this paper by its authors.
  Use permitted under Creative Commons License Attribution 4.0
  International (CC BY 4.0).}

\conference{To appear in AI for Access to Justice, Dispute Resolution, and Data Access Workshop(AI4A2J), 2025}

\title{LegalWebAgent: Empowering Access to Justice via LLM-Based Web Agents}

\author[1]{Jinzhe Tan}[
  email=jinzhe.tan@umontreal.ca,
]
\cormark[1]

\author[1]{Karim Benyekhlef}[]

\address[1]{Cyberjustice Laboratory, Faculty of Law, University of Montreal, Canada}

\cortext[1]{Corresponding author.}

\begin{abstract}
Access to justice remains a global challenge, with many citizens still finding it difficult to seek help from the justice system when facing legal issues. Although the internet provides abundant legal information and services, navigating complex websites, understanding legal terminology, and filling out procedural forms continue to pose barriers to accessing justice. This paper introduces the \texttt{LegalWebAgent} framework that employs a web agent powered by multimodal large language models to bridge the gap in access to justice for ordinary citizens. The framework combines the natural language understanding capabilities of large language models with multimodal perception, enabling a complete process from user query to concrete action. It operates in three stages: the Ask Module understands user needs through natural language processing; the Browse Module autonomously navigates webpages, interacts with page elements (including forms and calendars), and extracts information from HTML structures and webpage screenshots; the Act Module synthesizes information for users or performs direct actions like form completion and schedule booking. To evaluate its effectiveness, we designed a benchmark test covering 15 real-world tasks, simulating typical legal service processes relevant to Qu\'ebec civil law users, from problem identification to procedural operations. Evaluation results show \texttt{LegalWebAgent} achieved a peak success rate of 86.7\%, with an average of 84.4\% across all tested models, demonstrating high autonomy in complex real-world scenarios.
\end{abstract}


\maketitle

\section{Introduction}

The significant gap between the public's legal needs and their ability to find legal information and solutions has created a ``justice gap'' in countries worldwide~\cite{child2016global}. For ordinary citizens, the cost of hiring a lawyer is prohibitively expensive~\cite{hadfield2014cost}, forcing them to spend a great deal of time navigating a maze of government websites, legal statutes, and procedural forms on their own. Even so, they often find it difficult to search using correct legal terminology, locate relevant information on cluttered websites, or make critical errors when filling out online forms.

Existing legal tech tools, such as static frequently asked questions (FAQ) portals and simple rule-driven chatbots, seek to address this issue. They typically provide information in plain language\footnote{For example, Qu\'ebec's \'Educaloi (\url{https://educaloi.qc.ca/}) provides legal information to the public in plain language.}, or offer simplified legal pathways and relevant cases~\cite{westermann2023justicebot, janatian2023text} to help reduce the user's cognitive load. However, as the legal domains covered by these websites and the volume of information they contain increase, the cognitive burden on users increases correspondingly. Furthermore, these tools still lack the ability to interact with the broader web ecosystem or perform actions on behalf of the user. This means that the most difficult and error-prone practical ``operational'' steps, such as submitting forms or scheduling appointments, are still left to the user.

\begin{figure}[htbp]
  \centering
  \includegraphics[width=\textwidth]{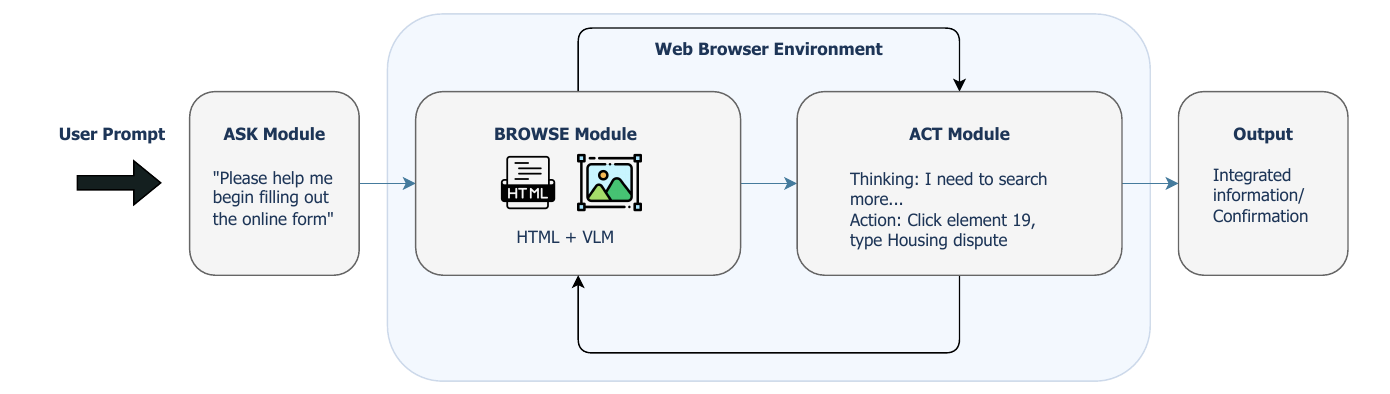}
  \caption{The overall workflow of \texttt{LegalWebAgent}. Given a user's query, \texttt{LegalWebAgent} formulates a plan, analyzes the webpage’s HTML elements and screenshots, and determines the appropriate actions (such as clicks, scrolls, or inputs). After gathering the necessary information or completing the requested task, it generates a concise summary of the process and presents the results to the user.}
  \label{fig:main}
\end{figure}

The recent development in the field of multimodal large language models (MLLMs) has advanced the role of AI in building a general-purpose tool to enhance access to justice. The improved reasoning capabilities of LLMs have made it possible to build autonomous agents~\cite{wang2024survey, ning2025survey}. \texttt{GPT-4o-vision} and more advanced LLMs demonstrate strong capabilities in natural language understanding and generation. Moreover, their multimodal understanding abilities allow for the simultaneous processing of inputs other than text, such as images~\cite{wu2023multimodal}, making it feasible to build web agents that can comprehend both the text and visual elements of webpages~\cite{zheng2024gpt}. In this study, we introduce the \texttt{LegalWebAgent} (see Figure~\ref{fig:main}) framework, a multimodal web agent designed to autonomously create plans based on a user's query to perform tasks such as web browsing, information gathering, and web interaction.

In the following sections, we review related work (Section~\ref{related_work}), introduce the \texttt{LegalWebAgent} framework (Section~\ref{framework}), outline our experiment design (Section~\ref{experiment}), and present results (Section~\ref{results}). We then discuss key insights and current limitations (Section~\ref{discussion}), and conclude.

\section{Related Work}\label{related_work}

\subsection{AI for Access to Justice}

The access to justice gap is a global issue that not only incurs monetary, temporal, and psychological costs~\cite{farrow2013access, currie2009legal, semple2015cost, farrow2016everyday, cannon2021unmet} but also erodes public confidence in the justice system. AI technology has been widely applied to bridge this gap~\cite{westermann2023using}, with efforts ranging from providing legal information~\cite{tan2023chatgpt, westermann2023bridging} and assisting with form completion~\cite{steenhuis2023beyond, westermann2024dallma} to online dispute resolution~\cite{branting2023computational, westermann2023llmediator, tan2024robots}. In addition to these primarily text-focused efforts, Vision-Large Language Models (VLLMs) have also undergone preliminary exploration in the legal domain in recent years~\cite{westermann2024analyzing}. Researchers have also explored integrating AI with information portals to help users map natural language descriptions to relevant legal issues~\cite{westermann2023bridging, steenhuis2024getting}. However, these systems are often built for specific contexts and thus lack generalizability. Furthermore, they are largely passive, requiring users to independently read, comprehend, and act upon the information provided.

\subsection{Web Agents}

The internet has long been one of the primary channels for users to obtain information. It can be viewed as a continuously updated real-time database. In the legal field, modifications to legal information, promulgation of new laws, and introduction of new cases are all published on the internet in a very timely manner. When needed, people can readily find the information they require online. Furthermore, tasks such as booking meetings, filling out online forms, and completing online questionnaires offer users the possibility of handling tasks remotely, saving them valuable time.

In practice, however, this process is far from straightforward: (1) users may encounter websites containing outdated or misleading information; (2) retrieving legal information often requires extensive cross-verification and synthesizing information from numerous web pages; (3) users may be unaware of the existence of certain authoritative resources; and (4) human cognitive limitations, along with states such as distraction and fatigue, can significantly hinder users’ ability to navigate the web effectively~\cite{wei2025browsecomp}.

The internet is designed for humans, enabling them to interact with the digital world through operations such as clicking, scrolling, and typing~\cite{gou2024navigating}. Creating web agents capable of simulating these operations is expected to significantly help reduce the sense of disorientation and inefficiency experienced during the aforementioned web browsing process.

Such web agents or web automation have been extensively explored for decades. Early research typically relied on structured data sources or site-specific wrappers to perform tasks such as flight booking or information scraping~\cite{hammer1997extracting, doorenbos1997scalable, kushmerick1997wrapper}. These methods required significant manual configuration for each website and were highly brittle, often breaking when sites changed~\cite{lerman2003wrapper, muslea1999hierarchical}.

With the development of deep learning and reinforcement learning, more \emph{generalized} web agents have become possible. Depending on the architecture, they range from agents trained purely on HTML files~\cite{shi2017world, liu2018reinforcement, xu2021grounding}, to models that combine HTML and visual screenshots~\cite{deng2023mind2web, lu2024weblinx}, and even to agents that rely solely on visual input from screenshots~\cite{chen2024guicourse}. These modern web agents are capable of helping users interact with real-world websites and perform everyday tasks~\cite{pahuja2025explorer}.

\section{\texttt{LegalWebAgent} Framework}\label{framework}

\begin{figure}[htbp]
  \centering
  \includegraphics[width=0.8\textwidth]{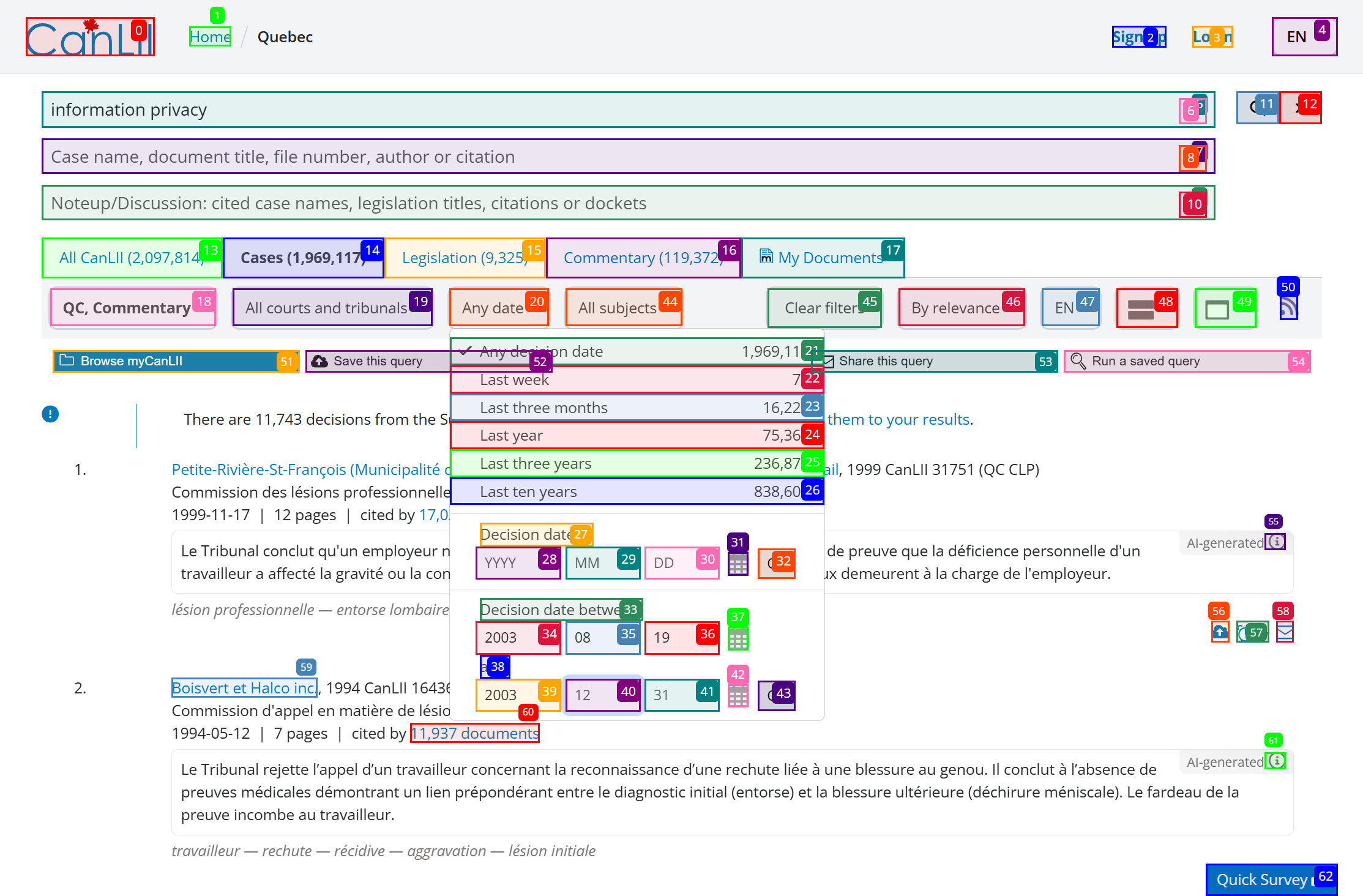}
  \caption{Example of a webpage screenshot provided to the web agent. By reading the HTML file, the application automatically adds borders to interactive elements on the webpage and labels them with numbered tags in the upper-right corner.}
  \label{fig:screenshot}
\end{figure}

\texttt{LegalWebAgent} is composed of three modules (Ask Module / Browse Module / Act Module) and works in concert with a web browsing environment. \texttt{LegalWebAgent} builds on top of the open-source \texttt{Browse-Use} framework~\cite{browser_use2024}, the core of which is \texttt{Playwright}~\cite{microsoft_playwright_2024}, which enables reliable and fast web automation by providing a unified API for Chromium, Firefox, and WebKit. When a user command is received, \texttt{LegalWebAgent} launches a Chromium browser and then provides context to the LLM by collecting the webpage's HTML elements and a screenshot (see Figure~\ref{fig:screenshot}). Based on the current state, \texttt{LegalWebAgent} analyzes the completion status of the previous objective, updates its memory, and formulates the next objective. According to the objective, \texttt{LegalWebAgent} proposes an action to be executed and performs the operation within the browser environment.

\subsection{Ask Module}

When the user provides a prompt (e.g., ``Find the rental dispute department closest to H3A0G4.''), the Ask module performs the following steps: (1) It uses an LLM to parse the user's intent. In this example, the module may recognize that the user needs legal information on landlord-tenant disputes in Qu\'ebec, as well as the address and phone number of the relevant tribunal. (2) Based on the parsed intent, the Ask module generates a web navigation plan. For instance, the plan might be:
\begin{quote}
1. Search for rental dispute resolution services;\\
2. The user's postal code is H3A0G4, focus the search on Montreal.
\end{quote}

\subsection{Browse Module}

The Browse module receives the current state of the webpage at each execution step and decides the next action to take. Because modern webpages often contain a large number of elements and complex layouts, we adopt a multimodal perception mechanism to improve the robustness of the Browse module.

\textbf{HTML Analysis.} The module parses the HTML Document Object Model (DOM) of the page, identifies candidate interactive elements (e.g., \texttt{<a>} links, \texttt{<button>} elements, form \texttt{<input>} fields), and extracts textual content. Using the method introduced in~\cite{yang2023set}, all candidate elements are automatically highlighted with bounding boxes, and a numbered tag is added in the upper-right corner of each box (see Figure~\ref{fig:screenshot}).

\textbf{Visual Analysis.} At the same time, the tagged page screenshot is passed to a VLLM. Since relying solely on DOM information can be misleading or incomplete (for instance, menus that appear only on hover), the visual input helps the LLM better handle different types of web pages.

After extracting webpage information, the Browse module must choose an action from a predefined action library. We define a set of primitive actions inspired by human browser interactions: \texttt{click(element)}, \texttt{input(text, field)}, \texttt{scroll(direction)}, \texttt{wait(seconds)}, etc.

\subsection{Act Module}

The Act module executes the actions provided by the Browse module in a real browser environment. If the action involves a click or text input, it performs the corresponding operation on the page and then returns control, along with the updated page state, back to the Browse module. This loop continues until either the predefined goal is achieved or the maximum number of execution steps is reached.

If the user requests an explanation or information, once the browsing stage has collected the relevant text, the Act module calls the LLM to generate a concise and user-friendly answer.

\begin{figure}[htbp]
  \centering
  \includegraphics[width=\textwidth]{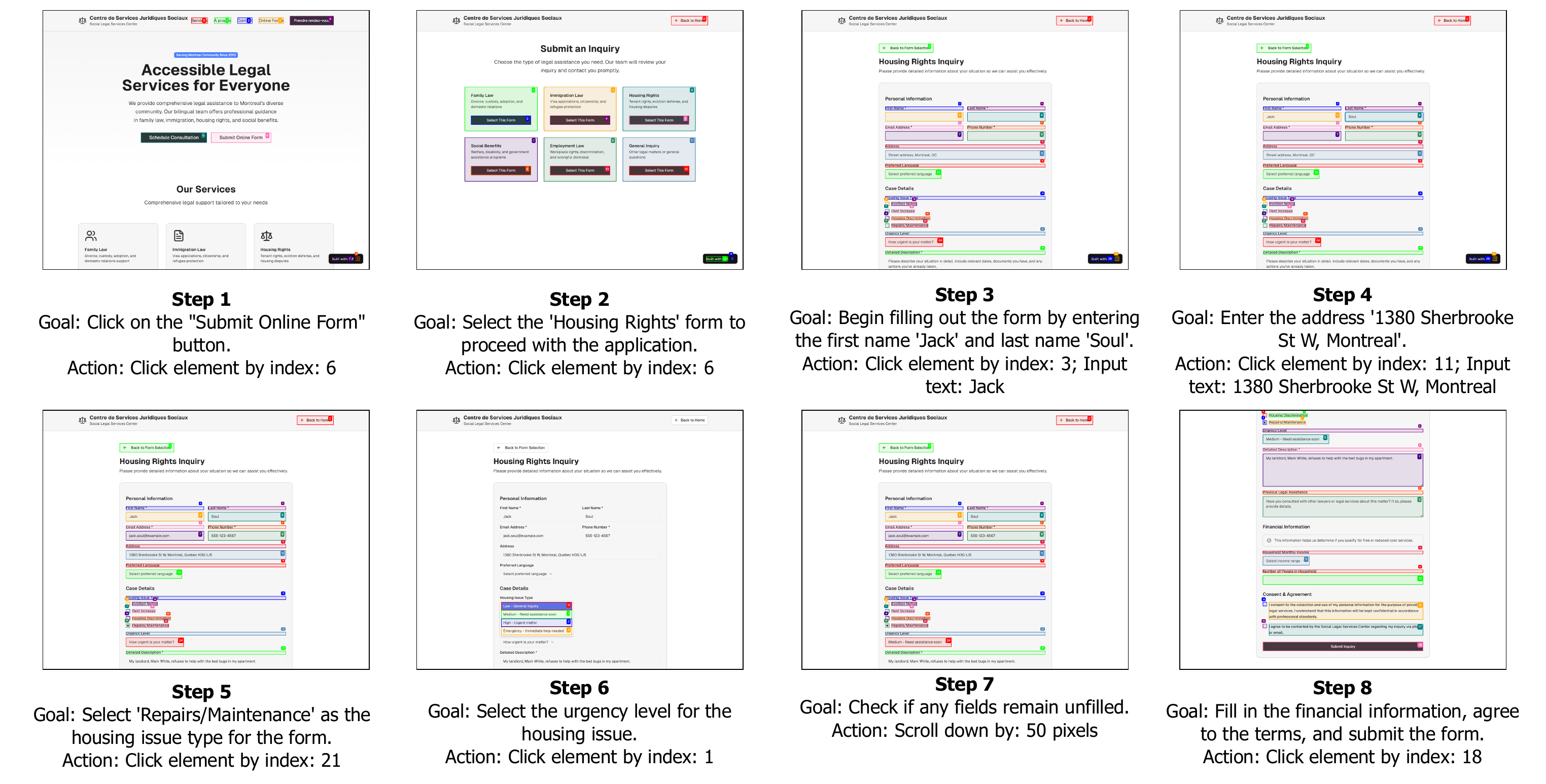}
  \caption{A demonstration of how \texttt{LegalWebAgent} navigates the website and completes an online form task through a series of actions on the \texttt{legal-agent-sandbox} we created.}
  \label{fig:steps}
\end{figure}

If the task objective involves performing an operation, the Act module verifies the completion of each sub-operation (e.g., filling in forms, uploading files, booking an appointment) and proceeds until the final submission. Upon completion, it outputs a confirmation message to the user (see Figure~\ref{fig:steps}), such as:
\begin{quote}
Form submitted successfully. Your confirmation number is 123-456.
\end{quote}

\section{Experiment Design}\label{experiment}

\subsection{Data Collection}

To evaluate the performance of \texttt{LegalWebAgent} on legal web tasks, we designed a preliminary benchmark suite comprising 15 tasks (described in Table~\ref{tab:task}). These tasks were selected from the domain of Qu\'ebec Civil Law and are designed to simulate real user queries or goals. The tasks are structured around three key stages of an ordinary citizen's legal journey:

\textbf{Information Gathering.} In this stage, the user is just starting to understand their legal problem. The tasks involve obtaining basic explanations of legal rights or concepts. For example, the user inquires about ``my landlord raised the rent without prior notice, saying I have to move out next month if I don't pay the new rent. Can he do that?'' This type of query primarily involves information retrieval and comprehension. \texttt{LegalWebAgent} must locate authoritative sources (such as government websites or legal information platforms) to find relevant resources and explain them to users in plain English.

\textbf{Resource Finding.} In this stage, the user is already aware of their general problem and is seeking specific resources or contact information. Task S2-LSA-01 asks, ``Where is the nearest rental dispute office to my postal code H3A0G4?'' This requires not only finding information about rental dispute offices but also locating the nearest one based on the user's location in Qu\'ebec. Task S2-CS-01 is, ``Search CanLii for a judgment date between August 19, 2023, and August 31, 2023, and a judgment length of exactly 23 pages.'' This task tests the agent's database retrieval capabilities. These tasks involve complex, multi-page information retrieval but do not include form submission.

\textbf{Action Taking.} This stage contains the most complex, multi-step tasks, requiring the agent to perform concrete actions on web pages. Task S3-OFC-01 is, ``Fill out the online form on \texttt{legal-agent-sandbox}.'' This simulates the scenario of a user submitting a claim or application through an official court web form. The agent must navigate a multi-page form interface, input predefined user information, interact with complex components (such as a calendar), and finally submit the application. These tasks significantly challenge the agent's autonomous navigation and form-handling capabilities, especially when dealing with dynamic elements such as date pickers or multi-step processes.

All tasks were executed on live or staged websites. For the first two stages, we require the agent to interact with actual public websites. For the third stage, which involves web form filling and online appointment scheduling, we built a \texttt{legal-agent-sandbox}\footnote{\url{https://legal-agent-sandbox.vercel.app/}} to avoid wasting public resources. This sandbox is deployed on a public domain and mimics the design of real websites, allowing us to test the agent's action-taking capabilities in a controlled environment.

\subsection{Models}

Given the high intelligence requirements of the web agent applications, we selected and evaluated three leading large language models: two vision-language models and one text-only model (to assess performance in an HTML-based plain-text context). We include one text-only model to measure how much performance depends on visual perception. Specifically, we tested OpenAI's \texttt{GPT-4o}, Anthropic's \texttt{Claude-Sonnet-4-20250514}, and DeepSeek's \texttt{DeepSeek-v3.1}.

\subsection{Experiments}

In the experiment, each model was given the same prompt and run at a temperature of 0.6. Once the prompt was sent to the model, no further intervention was made. After task completion, we scored whether the task was successful. A task was marked as successful if the model correctly fulfilled the request (e.g., providing accurate legal information, finding the correct institutions and locations, or submitting a form as instructed). It was marked as a failure if the model provided incorrect information, failed to retrieve useful information, or was unable to complete the required web interactions. We recorded the success/failure status, the number of reasoning/interaction steps, the time to completion, and the tokens consumed for each model on every task.

\begin{table}[htbp]
  \setlength{\tabcolsep}{4pt}
  \renewcommand{\arraystretch}{1.2}
  \footnotesize
  \centering
    \caption{\texttt{LegalWebAgent} benchmark overview.}
  \begin{tabularx}{\linewidth}{@{}P{0.12\linewidth}P{0.12\linewidth}cY@{}}
    \toprule
    \textbf{Stage} & \textbf{Category} & \textbf{Count} & \textbf{Query Description Example} \\
    \midrule
    \multirow{2}{*}{\makecell[tl]{Information\\Gathering}}
     & \makecell[tl]{Vague\\Inquiry} & 2 & User inquires about the legality of a landlord raising rent without prior notice and threatening eviction for non-payment of the new amount. \\
    \cline{2-4}
     & \makecell[tl]{Consumer\\Dispute} & 2 & User asks for recourse options after an expensive online purchase broke after one week and the seller refused to provide a refund. \\
    \midrule
    \multirow{3}{*}{\makecell[tl]{Resource\\Finding}}
     & \makecell[tl]{Complex\\Search} & 2 & User requests a search for a legal case on CanLii using highly specific criteria, including jurisdiction (Qu\'ebec), legal field, a narrow judgment date range, and exact document length. \\
    \cline{2-4}
     & \makecell[tl]{Locating\\Authority} & 3 & User needs to identify the correct government department for a rental dispute and find the location and phone number of the nearest office based on their postal code. \\
    \cline{2-4}
     & \makecell[tl]{Legal\\Aid} & 3 & User is looking for free or low-cost community legal services specializing in family law near downtown Montreal. \\
    \midrule
    \multirow{2}{*}{\makecell[tl]{Action\\Taking}}
     & \makecell[tl]{Form\\Completion} & 1 & User requests assistance in filling out an online legal application form with their personal details (name, address) and case description (landlord-tenant issue). \\
    \cline{2-4}
     & \makecell[tl]{Appointment\\Booking} & 2 & User instructs the system to schedule a legal consultation for them on a specific website for a given date. \\
    \bottomrule
  \end{tabularx}
  \label{tab:task}
\end{table}

\section{Results}\label{results}

\begin{figure}[htbp]
  \centering

  \begin{subfigure}[t]{0.48\textwidth}
    \centering
    \includegraphics[width=\linewidth]{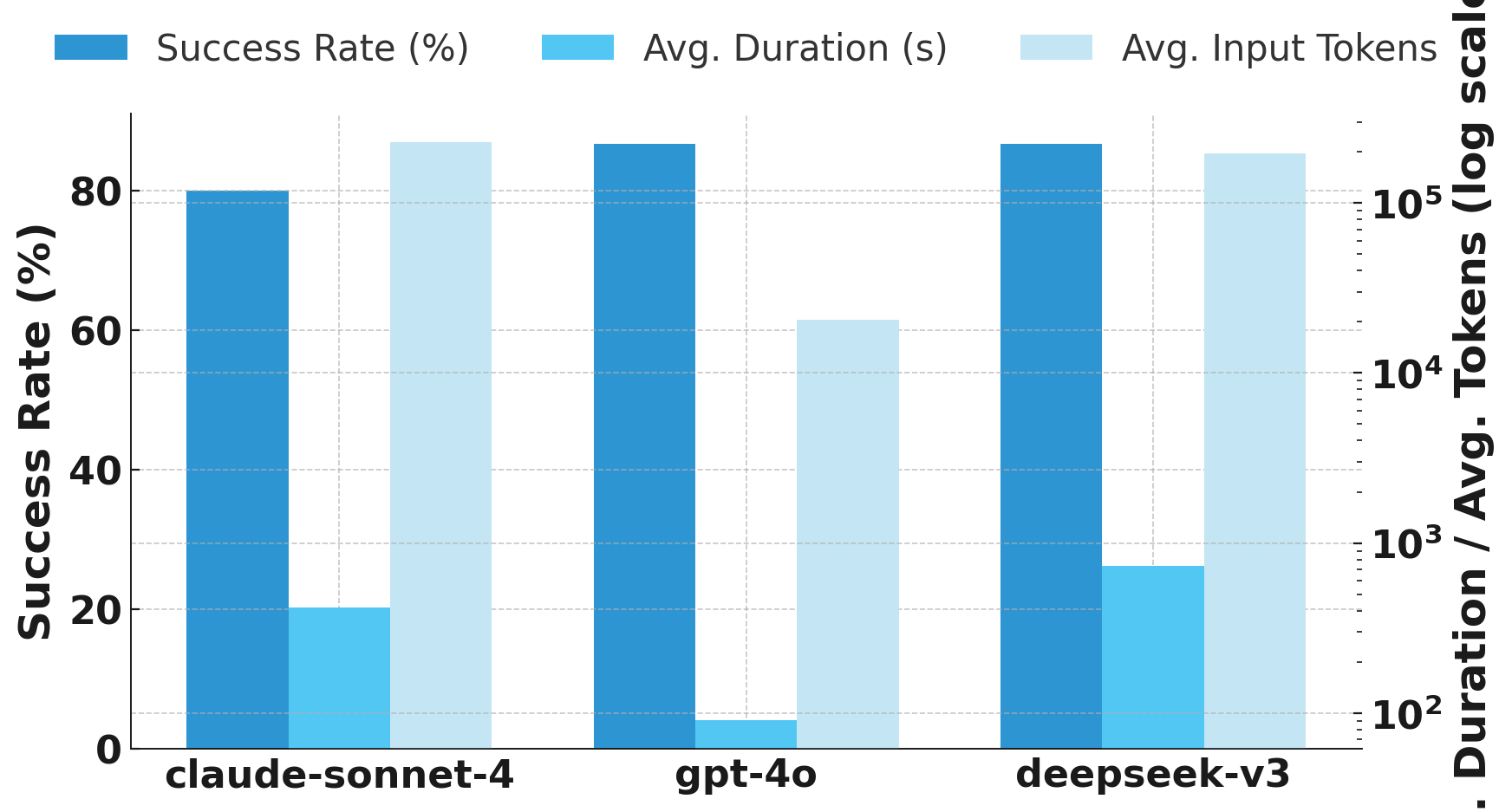}
    \subcaption{Overall performance comparison across models. The left y-axis shows the success rate (\%) for each model, while the right y-axis (log-scale) compares average duration (s) and average input tokens. \texttt{GPT-4o} achieves the highest efficiency (shortest duration, lowest token use), while \texttt{Claude-Sonnet-4} produces the most thorough outputs but at a higher computational cost.}
    \label{fig:success-rate}
  \end{subfigure}
  \hfill
  \begin{subfigure}[t]{0.48\textwidth}
    \centering
    \includegraphics[width=\linewidth]{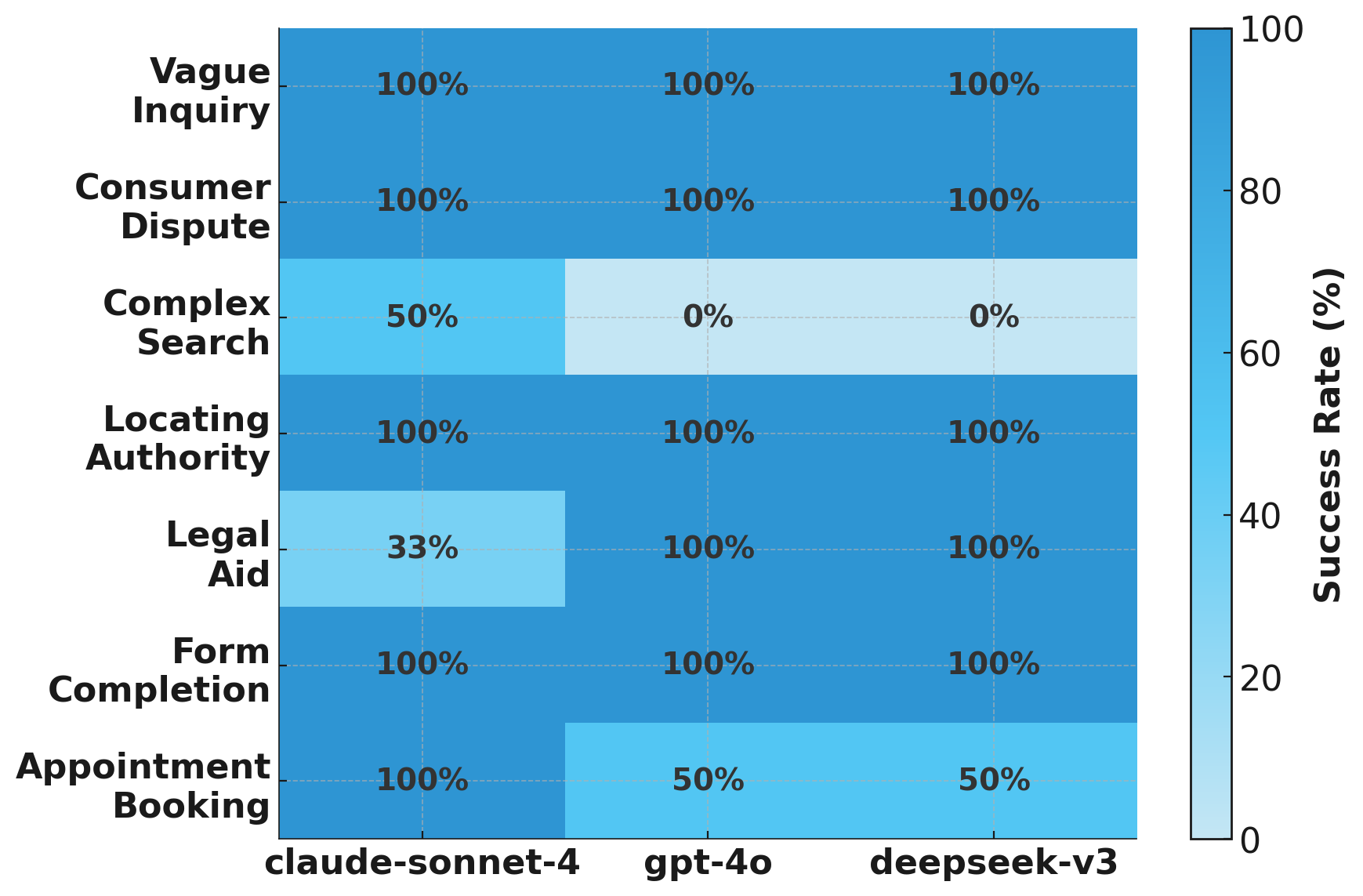}
    \subcaption{Heatmap of model success rates by task category. Darker cells indicate higher success rates. Most categories achieve 100\% success across all models except Complex Search (low performance across the board) and Appointment Booking, where \texttt{GPT-4o} and \texttt{DeepSeek-v3} underperform relative to \texttt{Claude-Sonnet-4}.}
    \label{fig:heatmap-cats}
  \end{subfigure}

  \caption{Overall performance comparison and task-wise success rates of the evaluated models.}
  \label{fig:heatmap}
\end{figure}

Table~\ref{tab:overall-compare-en} and Figure~\ref{fig:heatmap}~(a) show the models' overall task success and efficiency. OpenAI \texttt{GPT-4o} and \texttt{DeepSeek-v3} achieved the highest success rate at 86.7\% (13 out of 15 tasks each), while \texttt{Claude-Sonnet-4}, known for higher reasoning ability, completed 80.0\% (12/15 tasks). Notably, \texttt{GPT-4o} was by far the most efficient, with an average run time of only about 90 seconds per task and approximately 20k tokens consumed, compared to Claude's 416 seconds and 227k tokens. \texttt{DeepSeek-v3}'s runtime and token consumption are also an order of magnitude higher than \texttt{GPT-4o}'s (averaging 730 seconds and consuming 195k tokens). Claude-Sonnet-4 tends to generate extremely detailed and lengthy outputs while exploring more pages (resulting in high token consumption), yet this does not translate into higher success rates. In fact, Claude's comprehensive approach sometimes leads to diminishing returns, as it fails on tasks other models accomplish with fewer steps.

Figure~\ref{fig:heatmap}~(b) provides a heatmap of success rates across the seven task categories (see Table~\ref{tab:task} for task breakdown). We observe 100\% success across all models for most information-seeking tasks, including vague rights inquiries (S1-VRI) and consumer disputes (S1-CDD). All models also succeeded in locating authorities (S2-LSA) and completing the complex online form (S3-OFC) on our sandbox site. This suggests that current LLM agents are highly capable in straightforward information retrieval and form-filling scenarios in the legal domain.

The Complex Search (S2-CS) tasks had the lowest success rates: one CanLII case query (S2-CS-01), inspired by~\cite{wei2025browsecomp}, that is very difficult to search for but very easy to verify, was not solved by any model (0\% success). The other, a statistical query (S2-CS-02), was solved only by \texttt{Claude} (50\% success vs.\ 0\% for \texttt{GPT-4o} and \texttt{DeepSeek-v3} in this category). These complex searches require combining multiple filters or conditions on the website and involve extensive browsing to find the correct answer. Most failures were due to the models failing to correctly apply multiple filters simultaneously or abandoning the search prematurely.

Another challenging area was the online appointment booking system. In this task category, Claude-Sonnet-4 successfully completed both scheduling tasks (100\%), whereas \texttt{GPT-4o} and \texttt{DeepSeek-v3} each completed only one. Specifically, in task S3-OAB-02, which involves booking via an external Microsoft Forms page, both \texttt{GPT-4o} and \texttt{DeepSeek-v3} struggled. \texttt{DeepSeek-v3} failed after 50 repetitive steps on the booking interface, while \texttt{Claude} completed the form and secured the appointment in just 15 steps. This discrepancy suggests that despite its slower execution and higher cost, \texttt{Claude}'s more comprehensive planning strategy yields superior performance in handling highly complex, multi-step interactions.

\begin{table}[htbp]
  \footnotesize
  \centering
  \caption{Overall model comparison (success, efficiency, and input cost).}
  \begin{tabular}{@{}lcccc@{}}
    \toprule
    \textbf{Model} & \textbf{Success Rate} & \textbf{Successful Tasks (n)} & \textbf{Avg.\ Duration (s)} & \textbf{Avg.\ Tokens} \\
    \midrule
    \texttt{Claude-Sonnet-4} & 80.0\% & 12 & 416.32 & 227{,}594 \\
    \texttt{GPT-4o}          & 86.7\% & 13 & 90.9   & 20{,}514  \\
    \texttt{DeepSeek-v3}     & 86.7\% & 13 & 730.0  & 195{,}519 \\
    \bottomrule
  \end{tabular}

  \label{tab:overall-compare-en}
\end{table}

\section{Discussion}\label{discussion}

Our initial evaluation suggests that web agents powered by large language models can autonomously perform information gathering, resource discovery, and take action to promote access to justice. \texttt{LegalWebAgent} demonstrated strong performance on our tasks, achieving high success rates in answering legal questions, locating government agencies, and even completing online form submissions. However, we also observed challenges during experimentation that would arise when deploying such agents in real-world scenarios.

\subsection{Deep or Success?}

We observed that tasks requiring deep website navigation (many clicks or layers of pages) had a higher failure probability. Advanced models like \texttt{Claude-Sonnet-4} tended to ``dig deeper'' into sites, which sometimes helped but often led to getting sidetracked or timing out. Excessive exploration can cause the agent to lose context or encounter dead ends (e.g., endlessly following related links). An effective agent must balance thoroughness with focus. Future agents might benefit from adaptive search depth control, stopping when sufficient information is found rather than exhaustively clicking every link.

\subsection{Web Design for AI Agents}

Our findings suggest that today’s websites, designed for human users, can inadvertently confuse or hinder AI agents. For example, pop-ups like cookie consent or privacy policy notices are easy for humans to handle but can disrupt an agent’s understanding. Similarly, information hidden within complex menus or behind interactive components is difficult for web agents to extract. To improve agent performance (and perhaps human usability too), web developers could consider offering ``AI-friendly'' design, such as simplified HTML views, Markdown or JSON endpoints, or well-labeled content, tailored for autonomous agents.

\subsection{Token Efficiency}

The difference in token consumption across different models for the same task highlights the importance of optimizing how agents process web content. \texttt{Claude-Sonnet-4} tends to load entire pages or even multiple pages in detail, leading to an exceptionally large context window. This not only impacts agent execution speed but may also trigger LLM context-length limitations. We can reduce token consumption by optimizing the content extracted by agents, which is important for cost reduction and efficiency improvement in web agents as well as their real-world deployment.

\subsection{Limitations}

This research is a preliminary exploration of web agents for promoting access to justice, tested on a limited dataset (15 tasks) focused on Qu\'ebec’s legal context. While our dataset was designed to reflect genuine user needs, broader evaluation is required for more accurate results. Future work should expand the testing scope to include more diverse legal scenarios, such as different jurisdictions, various legal websites, and a wider array of legal problem types.

A further limitation is the framework's dependence on large proprietary language models, which may incur excessive operational costs. Investigating smaller or open-source models and enhancing agent efficiency are therefore worthwhile research avenues. Agent safety is another important concern. For instance,~\cite{kumar2024refusal} suggests web agents might be more vulnerable to jailbreaking, and~\cite{zychlinski2025whole} discusses how developers could add a ``toxic'' page, visible only to agents, to manipulate their behavior. This indicates that the present study is experimental in nature and requires critical robustness testing and security reviews before practical deployment.

\section{Conclusion}

The \texttt{LegalWebAgent} framework showcases the potential of multimodal LLM-based agents in bridging the ``justice gap.'' By combining MLLMs with web navigation and action execution, such agents can empower individuals to find legal information and complete online procedures with minimal manual effort. Our model comparison results indicate that while current state-of-the-art LLMs can reliably handle most tasks, they still require fine-tuning to effectively manage complex searches and intricate web forms. There remains significant room for improvement in agent response speed, reliability, and web content parsing. We believe that by automating web browsing processes, web agents based on MLLMs have strong potential to enhance access to justice.

\section*{Declaration on Generative AI}

During the preparation of this work, the authors used large language model tools (such as GPT-style assistants) for grammar and spelling checking, wording suggestions, and minor text refinement. After using these tools, the authors reviewed and edited the content as needed and take full responsibility for the publication’s content.


\bibliography{biblio}

\appendix



\end{document}